\documentclass[aps,showpacs,showkeys,preprint]{revtex4}
\usepackage[dvips]{graphicx}
\usepackage{latexsym}
\begin{document}

\title{Ruppeiner geometry and 2D dilaton gravity in the thermodynamics of black holes}

\author{Yun Soo Myung\footnote{ysmyung@inje.ac.kr}}
\affiliation{Institute of Basic Science and
  School of Computer Aided Science, Inje University, Gimhae 621-749,
  Korea}
\author{Yong-Wan Kim\footnote{ywkim65@gmail.com}}
\affiliation{Institute of Basic Science and
  School of Computer Aided Science, Inje University, Gimhae 621-749,
  Korea}
\author{Young-Jai Park\footnote{yjpark@sogang.ac.kr}}
\affiliation{Department of Physics  and Center for Quantum
  Spacetime, Sogang University, Seoul 121-742, Korea}

\begin{flushright}
\today
\end{flushright}

\begin{abstract}

We study the geometric approach to the black hole thermodynamics.
The geometric description of the equilibrium thermodynamics comes
from Ruppeiner geometry based on a metric on the thermodynamic
state space. For this purpose, we consider the
Reissner-Nordstr\"om-AdS (RN-AdS) black hole which provides two
different ensembles: canonical ensemble for fixed-charge case and
grand canonical ensemble for fixed-potential case. Two cases are
independent and cannot be mixed into each other.  Hence, we
calculate different Ruppeiner curvatures for two ensembles.
However, we could not find the consistent behaviors of Ruppeiner
curvature corresponding to those of heat capacity. Alternatively,
we  propose the curvature scalar in the 2D dilaton gravity
approach which shows the features of  extremal, Davies and minimum
temperature points of RN-AdS black hole, clearly.
\end{abstract}


\maketitle

\section{Introduction}

Since the pioneering work of Bekenstein \cite{bek,bek2,bek3} and
Hawking \cite{haw}, which proved that the entropy of black hole is
proportional to the area of its horizon in the early 1970s, the
research of the black hole  thermodynamics has greatly improved.
Especially, it was found that if the surface gravity of the black
hole is considered to be the temperature and the outer horizon
area is considered to be the entropy, four laws of the black hole
thermodynamics, which correspond with the four laws of the
elementary thermodynamics, has been established~\cite{BCH}.
Recently, there are several geometric approaches~\cite{wr} to the
black hole thermodynamics~\cite{dav,rup,abp,que,SCWS,mz}. The
geometric description of the equilibrium thermodynamics comes from
Ruppeiner geometry based on the metric of the thermodynamic state
space: entropy metric and others including Weinhold
metric~\cite{wein}. However, the results depends on how to choose
the  metric. Very recently, Ruppeiner has discussed on  a
systematic discussion of how to make the correct choice of a
metric~\cite{Rup1}.  Also, Ruppeiner has demonstrated several
limiting results matching extremal Kerr-Newman black hole
thermodynamics to the two dimensional Fermi gas~\cite{Rup2},
showing that this connection to a 2D model is consistent with the
membrane paradigm of black holes~\cite{Tho}.

On the other hand, 2D dilaton gravity has been used in various
situations as an effective description of 4D gravity after a black
hole in string theory has appeared \cite{witten,wit1}. In
particular, thermodynamics of this black hole has been analyzed by
several authors \cite{crff,crf1,crf2,gkv}. Another 2D theories,
which were originated from the Jackiw-Teitelboim (JT) theory
\cite{jackiw,Teit}, have been also studied
\cite{JT-theories,JT2,fnn}. Recently, we have introduced the 2D
dilaton gravity approach, which completely preserves the
thermodynamics of 4D black hole~\cite{mod}. We have obtained that
the 2D curvature scalar shows the features of  extremal, Davies
and minimum temperature points of RN black hole, clearly.
Actually, since the heat capacity is inversely proportional to the
2D curvature scalar, we expect that the 2D curvature could show
interesting  thermodynamic points clearly.

In this paper, we  deal with the issue of geometric approach by
analyzing the RN-AdS black hole, which provides two different
ensembles: canonical ensemble for fixed-charge case and grand
canonical ensemble for fixed-potential case~\cite{CEJM}. Since
these are independent of each other, one cannot mix them by
Legendre transformation. Hence, we have to calculate two Ruppeiner
curvatures for  different ensembles. In Sec. II, we briefly
recapitulate the RN-AdS black hole thermodynamics. In Sec. III, we
calculate different Ruppeiner curvatures based on fixed-charge and
fixed-potential ensembles. In Sec. IV, alternatively, we propose
the 2D curvature scalar, which shows the features of  extremal,
Davies and minimum temperature points of the RN-AdS black hole,
clearly. In Sec. V, we point out the different thermodynamic
behaviors between Ruppeiner geometry and 2D dilaton gravity in the
thermodynamics of black holes. Finally, discussions are devoted to
Sec. VI.

\section{ RN-AdS black hole thermodynamics}
We start with the four-dimensional action
\begin{equation}
\label{action} I_4=\frac{1}{16\pi G}\int d^4x
          \sqrt{-g}\Bigg[R_4-F_{\mu\nu}F^{\mu\nu}+\frac{6}{l^2}\Bigg]
\end{equation}
with $l$ the curvature radius
of AdS$_4$ spacetimes. Hereafter we shall adopt Planck units of
$G=\hbar=c=k=1$. The solutions to equations of motion lead to the
RN-AdS black hole whose metric is given by
\begin{equation} \label{MF}
ds^2_{RN-AdS}=-U(r)dt^2+U^{-1}(r)dr^2+r^2d\Omega^2_2
\end{equation}
with the metric function $U(r)$
\begin{equation} \label{adsrnMF}
U(r)=1-\frac{2M}{r}+\frac{Q^2}{r^2}+\frac{r^2}{l^2},
\end{equation}
and $F_{\mu\nu}F^{\mu\nu}=-2Q^2/r^4$. Here, $d\Omega^2_2$ denotes
$d\theta^2+\sin^2\theta d\varphi^2$. We note that the reduced mass
becomes the ADM mass ($m=2M$) and the charge parameter becomes the
charge ($q=Q$) when choosing Planck units.
 Then, the inner ($r_-$) and the outer ($r_+$) horizons are
obtained from the condition of  $U(r_\pm)=0$. Using these
horizons, the mass and charge are expressed  by
\begin{equation}
M(r_+,r_-)=\frac{1}{2}\Bigg[r_++r_-+
\frac{r_+^4-r_-^4}{l^2(r_+-r_-)}\Bigg],~Q^2(r_+,r_-)=r_+r_-\Bigg[1+\frac{r_+^3-r_-^3}{l^2(r_+-r_-)}\Bigg].
\end{equation}

First, let us consider the case of fixed-charge ensemble.
For the RN-AdS black hole~\cite{CEJM}, the relevant thermodynamic
quantities are given by the Bekenstein-Hawking entropy $S$, ADM
mass $M_Q$, and Hawking temperature $T_Q$
\begin{equation} \label{aas}
S(r_+)=\pi r_+^2,~M_Q(r_+,Q)=\frac{r_+}{2}\Bigg[1+\frac{Q^2}{r_+^2}+\frac{r_+^2}{l^2}\Bigg],~
T_Q(r_+,Q)= \frac{1}{4\pi}\Bigg[
\frac{1}{r_+}-\frac{Q^2}{r_+^3}+\frac{3r_+}{l^2}\Bigg].
\end{equation}
In this case that the horizon is degenerate ($r_+=r_-=r_e$), we
have an extremal black hole with $M=M_e=r_e$. In general, one has
an inequality of $M>M_e$. Then, using the Eq. (\ref{aas}), the
heat capacity $C_Q=(dM_Q/dT_{Q})_Q$ for fixed-charge $Q$ takes the
form
\begin{equation}\label{aac}
C_Q(r_+,Q)= 2\pi r_+^2
\Big[\frac{3r_+^4+l^2(r_+^2-Q^2)}{3r_+^4+l^2(-r_+^2+3Q^2)}\Big].
\end{equation}
The global features of thermodynamic quantities are shown in Fig. 1 for
$Q<Q_c=l/6$.  Here we observe the local minimum
$T_Q=T_0$ at $r_+=r_{0}$ (Schwarzschild-AdS black hole), in addition to the zero temperature $T_H=0$ at
$r_+=r_e$(extremal RN black hole) and the local maximum  $T_Q=T_D$
at $r_+=r_{D}$ (Davies' point of RN black hole).
It seems to be a combination of the RN and
Schwarzschild-AdS black holes.

Here, we observe that $C_Q=0$ and $T_Q=0$ at $r_+=r_e$,
\begin{equation}
r_e^2=\frac{l^2}{6}\Bigg[-1+\sqrt{1+\frac{12Q^2}{l^2}}\Bigg].
\end{equation}
and  the heat capacity  blows up at $r_+=r_D$ and $r_0$ where
these satisfy
\begin{equation}
r_D^2=\frac{l^2}{6}\Bigg[1-\sqrt{1-\frac{36Q^2}{l^2}}\Bigg],
~r_0^2=\frac{l^2}{6}\Bigg[1+\sqrt{1-\frac{36Q^2}{l^2}}\Bigg].
\end{equation}
These points exist only for $Q<Q_c= l/6$. For $Q=Q_c$, we have
$r_D=r_0=l/\sqrt{6}$. Here we consider the thermal stability of a
black hole. The local stability is usually determined  by the
positive sign of heat capacity by considering evaporation and
absorbing processes of a black hole, even for an exceptional case
of the Kerr-Newman black hole~\cite{TL}. For example, the heat
capacity of the Schwarzschild black hole is $-2\pi r_+^2$, which
means that this isolated black hole is not in equilibrium in
asymptotically flat spacetimes.  Based on the local stability of
heat capacity, the RN-AdS black holes of $Q<Q_c$ can be split into
small stable black hole with $C_Q>0$ being in the region of
$r_e<r_+<r_D$, intermediate unstable black hole with $C_Q<0$ in
the region of $r_D<r_+<r_0$, and large stable black hole with
$C_Q>0$ in the region of $r_+>r_0$.
\begin{figure*}[t!]
   \centering
   \includegraphics{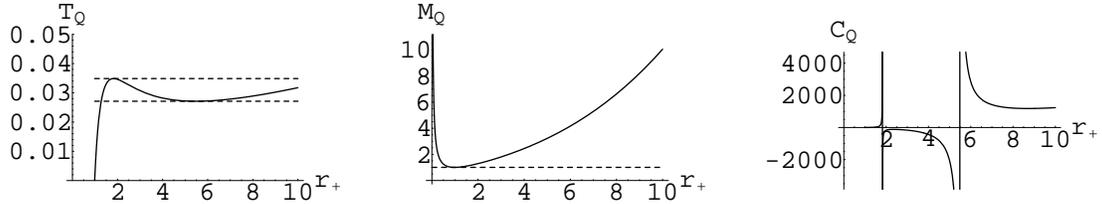}
\caption{Thermodynamic quantities of the RN-AdS black hole as
function of horizon radius $r_+$ with fixed $Q=1<Q_c$ and $l=10$:
temperature $T_Q$ with $T=T_D=0.035,T_0=0.027$(dashed lines), mass
$M_Q$ with $M=M_e=1.005$ (dashed line) and heat capacity $C_Q$. }
\end{figure*}

On the other hand, for fixed-potential $\Phi=Q/r_+$ ensemble~\cite{CEJM},
we have the mass and Hawking temperature
\begin{equation} \label{aasp}
M_\Phi(r_+,\Phi)=\frac{r_+}{2}\Bigg[1+\Phi^2+\frac{r_+^2}{l^2}\Bigg],~
T_{\Phi}(r_+,\Phi)= \frac{1}{4\pi r_+}\Bigg[
1-\Phi^2+\frac{3r_+^2}{l^2}\Bigg].
\end{equation}
In this ensemble, a relevant thermodynamic quantity is the
internal energy $J_\Phi$ defined by
\begin{equation} \label{aasie}
J_\Phi(r_+,\Phi)=M_\Phi-\Phi
Q=\frac{r_+}{2}\Bigg[1-\Phi^2+\frac{r_+^2}{l^2}\Bigg].
\end{equation}
Furthermore, the heat capacity $C_\Phi=T_\Phi(\partial S/\partial
T_\Phi)_\Phi$ takes the form
\begin{equation}\label{aacp}
C_\Phi(r_+,\Phi)= 2\pi r_+^2
\Big[\frac{3r_+^2+l^2(1-\Phi^2)}{3r_+^2+l^2(-1+3\Phi^2)}\Big].
\end{equation}
In this grand canonical ensemble, the points $r_+=r_e,~r_D,~r_0$ of
zero-temperature and the blow-up heat capacity are not those
in the canonical ensemble~\cite{Wu}.
Furthermore, if we compare Fig. 2 for $Q<Q_c$
with  Fig. 3 for $\Phi<\Phi_c$, we could find that
$C_\Phi$ displays no singular behavior
at the singular points $r_+=r_D,~r_0$ of $C_Q$.
Hence we must separate the fixed-potential case from the fixed-charge one.

\section{Ruppeiner curvature}
\subsection{($S,Q$)-representation}

Another observation on the geometric description of the
equilibrium thermodynamics comes from Ruppeiner geometry based on
a metric on the thermodynamic state space~\cite{rup}
\begin{equation}
ds_R^2=g_{ij}^Rdx^idx^j=\frac{1}{T_Q}g_{ij}^Wd\tilde{x}^id\tilde{x}^j,~x^i=(M,Q),~\tilde{x}^i=(S,Q),
\end{equation}
where the Ruppeiner metric $g_{ij}^R$ and the Weinhold metric
$g_{ij}^W$ are given by
\begin{equation}
g_{ij}^R=-\frac{\partial^2 S(x^i)}{\partial x^i \partial
x^j},~g_{ij}^W=\frac{\partial^2 M_Q(\tilde{x}^i)}{\partial
\tilde{x}^i
\partial \tilde{x}^j}.
\end{equation}
In this case, we use the Ruppeiner metric as
\begin{equation}
g^Q_{ij}=\frac{1}{T_Q}\frac{\partial^2 M_Q} {\partial \tilde{x}^i
\partial \tilde{x}^j}
\end{equation}
whose  form is given by
\begin{eqnarray}
g^Q_{ij}=\frac{1}{2S(1-\frac{\pi Q^2}{S}+\frac{3S}{\pi
l^2})}\left(
\begin{array}{cc} \label{metQ}
-1+\frac{3\pi Q^2}{S}+\frac{3S}{\pi l^2}& 4\pi Q \\
4\pi Q & 8\pi S
\end{array}
\right)
\end{eqnarray}

We observe the important relation
\begin{equation}
g^Q_{SS}=\frac{1}{T_Q} \Big(\frac{\partial^2 M_Q}{\partial
S^2}\Big)_Q=\frac{1}{C_Q(S,Q)}.
\end{equation}
Its determinant takes the form
\begin{equation}
{\rm Det}[g_{ij}]= - \frac{2\pi}{S} \frac{(1-\frac{\pi
Q^2}{S}-\frac{3S}{\pi l^2})}{(1-\frac{\pi Q^2}{S}+\frac{3S}{\pi
l^2})^2}
\end{equation}
One Ruppeiner curvature is given by~\cite{abp}
\begin{equation}
{\cal R}^I_{Q} = \frac{9}{\pi l^2}\frac{(1-\frac{\pi Q^2}{S}-\frac{S}{\pi
l^2})(\frac{\pi Q^2}{S}+\frac{3S}{\pi l^2})}{(1-\frac{\pi
Q^2}{S}+\frac{3S}{\pi l^2})(1-\frac{\pi Q^2}{S}-\frac{3S}{\pi
l^2})^2}.
\end{equation}
In this work, we use the computer program {\it Mathematica} 6.0
and {\it General Relativity, Einstein \& All} package (GREAT.m)
\cite{great} for the calculations of all Ruppeiner curvatures. We
observe that the curvature diverges in the extremal limit of
$1-\frac{\pi Q^2}{S}+\frac{3S}{\pi l^2}=0(r_+=r_-)$. However, the
two blow-up points of heat capacity which satisfy $1-\frac{3\pi
Q^2}{S}-\frac{3S}{\pi l^2}=0(3r_+^4-l^2 r^2_+ + 3l^2Q^2=0)$
disappear in the denominator of $R^I_{Q}$. This arises because
off-diagonal terms in $g^Q_{ij}$ contribute to calculating
Det[$g_{ij}$]. Dimensional reduction from KN-AdS black hole (when
$J\to0$) provides the other Ruppeiner curvature~\cite{mz}
\begin{equation}
{\cal R}^{II}_{Q} \propto - \frac{1}{(\pi l^2)^7 S^6 (1+\frac{S}{\pi
l^2})^2 \Big(1+\frac{\pi Q^2}{S}+\frac{S}{\pi l^2}\Big)^2
\Big(1-\frac{\pi Q^2}{S}+\frac{3S}{\pi l^2}\Big)\Big(1-\frac{\pi
Q^2}{S}-\frac{3S}{\pi l^2}\Big)^2 },
\end{equation}
which shows that this curvature diverges at the extremal point but
is finite at $r_+=r_D,~r_0$.

\subsection{($S,\Phi$)-representation}
In this case, we use the Ruppeiner metric as
\begin{equation}
g^\Phi_{ij}=\frac{1}{T_\Phi}\frac{\partial^2 J_\Phi(S,\Phi)}
{\partial \tilde{x}^i
\partial \tilde{x}^j}
\end{equation}
whose  form is given by
\begin{eqnarray} \label{metP}
g^\Phi_{ij}=\frac{1}{2S(1-\Phi^2+\frac{3S}{\pi l^2})}\left(
\begin{array}{cc}
-1+\Phi^2+\frac{3S}{\pi l^2}& -4\Phi S \\
-4\Phi S & -8S^2
\end{array}
\right).
\end{eqnarray}
Here, we use $J_\Phi=M_\Phi-\Phi Q$ instead of $M_\Phi$ because
the former could be defined  in the potential
representation~\cite{SCWS}. Here, we call $J_{\Phi}$ as a free
energy since the Gibbs potential is defined by $G=J_{\Phi}-T_\Phi
S$ in the grand canonical ensemble.

We note that $T_Q=T_\Phi$, but they have different
representations. We observe the important relation
\begin{equation}
g^\Phi_{SS}=\frac{1}{T_\Phi} \Big(\frac{\partial^2
J_\Phi}{\partial S^2}\Big)_\Phi=\frac{1}{C_\Phi(S,\Phi)}.
\end{equation}
 Its determinant takes the form
\begin{equation}
{\rm Det}[g_{ij}]=2 \left(\frac{1-3\Phi^2-\frac{3S}{\pi
l^2}}{1-\Phi^2+\frac{3S}{\pi l^2}}\right).
\end{equation}
Then, Ruppeiner curvature is given by
\begin{equation}
{\cal R}_{\Phi} =
-\frac{(1-\Phi^2)^2-\frac{18S^2}{\pi^2l^4}(1-3\Phi^2)+\frac{3S}{\pi^2l^4}(3-9\Phi^2+10\Phi^4)}{S(1-\Phi^2+\frac{3S}{\pi
l^2})(1-3\Phi^2-\frac{3S}{\pi l^2})^2}.
\end{equation}
We observe that the curvature diverges at the extremal black hole
of  $S=S_e$ which satisfies $1-\Phi^2+\frac{3S}{\pi l^2}=0$,
existing for $\Phi \ge \Phi_c$.

Finally, we comment that it is meaningless to get on-diagonal
form by starting from $g^Q_{ij}$ in Eq. (\ref{metQ}) and then
conformally transforming with a new coordinate of $u=\Phi$~\cite{abp}. The
resulting metric is just the diagonal part of $g^\Phi_{ij}$ in
Eq. (\ref{metP}).  This is because it gives rise to mixing between
fixed-charge  $g^Q_{ij}$ and fixed-potential $g^\Phi_{ij}$.

\section{2D dilaton gravity induced by dimensional reduction}
Assuming ${\cal M}_4={\cal M}_2 \times S^2$ for our purpose, we
perform a Kaluza-Klein reduction
\begin{equation}\label{non-conf}
ds^2_{RN-AdS}=h_{\alpha\beta}(t,r)dx^{\alpha}dx^{\beta}+b^2d\Omega^2_2,
\end{equation}
where $b$ represents the radius of two sphere $S^2$. After the
dimensional reduction by integrating Eq.(\ref{action}) over $S^2$,
the 2D  action is given by
\begin{equation} \label{2dact}
I_2=\frac{1}{4}\int_{{\cal M}_2} drdt\sqrt{-h}
     \Bigg[b^2 R_2(h)+2h^{\alpha\beta}\nabla_\alpha b\nabla_\beta b+2-
     b^2\Bigg(F_{\alpha\beta}F^{\alpha\beta}-\frac{6}{l^2}\Bigg)\Bigg].
\end{equation}
Here $R_2(h)$ is the 2D Ricci scalar. For thermodynamic analysis,
it is convenient to eliminate the kinetic term by using the
conformal transformation~\cite{GKL,cfnn,mkp}
\begin{equation} \label{conft}
\bar{h}_{\alpha\beta}=\sqrt{\phi}~h_{\alpha\beta},~\phi=\frac{b^2}{4}.
\end{equation}
The 2D  action is given by
\begin{equation} \label{2dactt}
\bar{I}_2=\int_{\bar{{\cal M}}_2} dxdt \sqrt{-\bar{h}}\Bigg[\phi
\bar{R}_2+\frac{1}{2\sqrt{\phi}}-\phi^{3/2}\bar{F}_{\alpha\beta}\bar{F}^{\alpha\beta}
+\frac{6\sqrt{\phi}}{l^2}\Bigg].
\end{equation}
where  $\bar{R}$ is the 2D Ricci scalar in the conformal
transformed frame. This transformation delivers information on the
4D action (\ref{action}) to the 2D dilaton potential completely,
if the 4D action provides the black hole solution~\cite{mod}.
That is, we may get the good $s$-wave approximation to the 4D
black hole without kinetic term for thermodynamic analysis. Unless
one makes the conformal transformation, the information is split
into the kinetic and the potential terms~\cite{GM}.

Making use of equation of motion for the
electromagnetic field yields to~\cite{NSN}
\begin{equation}
\frac{\phi^{3/2}\bar{F}_{+-}}{\sqrt{-\bar{h}}}=\bar{Q},
\end{equation}
where $\bar{Q}$ is an integration constant. In order to recover
the RN-AdS black hole, we choose $\bar{Q}=Q/4$ so that
\begin{equation}
\bar{F}_{\alpha\beta}\bar{F}^{\alpha\beta}=2\bar{F}_{+-}^2=\frac{Q^2}{8\phi^3}.
\end{equation}
  The 2D effective action is
obtained as
\begin{eqnarray}\label{repara-action}
\bar{I}_2=\int_{\bar{{\cal M}}_2} dxdt
\sqrt{-\bar{h}}[\phi\bar{R}_2+V_Q].
\end{eqnarray}
For the fixed charge ensemble, the dilaton potential is given by
\begin{equation}
V_Q(\phi,Q)=\frac{1}{2\sqrt{\phi}}-\frac{Q^2}{8\phi^{3/2}}
+\frac{6\sqrt{\phi}}{l^2}. \label{pot}
\end{equation}
This is an effective 2D dilaton gravity. The two equations of
motion are
\begin{eqnarray} \label{newat1}
\bar{\nabla}^2\phi=V_Q(\phi),\\
\bar{R}_Q=-\frac{dV_Q}{d\phi},  \label{newat2}
\end{eqnarray}
where the derivative of $V$ is given by
\begin{equation} \label{derivv}
\frac{dV_Q}{d\phi}=-\frac{1}{4\phi^{3/2}}+\frac{3Q^2}{16\phi^{5/2}}+\frac{3}{l^2\sqrt{\phi}}.
\end{equation}
The former is called  the dilaton equation and the latter is the
Einstein equation, even though the former(latter) are obtained
from metric(dilaton) variations. In order to solve the above
equations, we introduce the Schwarzschild-type  metric for
$\bar{h}_{\alpha\beta}$ as
\begin{equation} \label{2dmetric}
\bar{h}_{\alpha \beta}={\rm diag}(-f,f^{-1}).
\end{equation}
Then, its curvature scalar takes the form
\begin{equation} \label{2dcur}
\bar{R}_Q=-f'',
\end{equation}
where the prime ``$\prime$" denotes
the derivative with respect to $x$.
We note  from Eq.(\ref{conft}) that the dilaton is independent of time
$t$($\phi=\phi(x)$). Then, Eqs. (\ref{newat1}) and (\ref{newat2}) reduce to
\begin{equation} \label{eqp}
f\phi''+f'\phi'=V_Q(\phi),~f''=\frac{dV_Q}{d\phi}.
\end{equation}
In addition, we have the kinetic term for $\phi$ \begin{equation}
\label{phisq} (\bar{\nabla}\phi)^2=f(\phi')^2.
\end{equation}
If one chooses the linear dilaton as the solution
\begin{equation}
\phi=x,
\end{equation}
then Eq.(\ref{eqp}) leads to
\begin{equation}
f'=V_Q,~f''=V_Q',
\end{equation}
which implies that the latter is just a redundant relation.
We obtain the solution to Eqs. (\ref{newat1}) and (\ref{newat2})
\begin{equation}
 ds^2_{2D}=\bar{h}_{\alpha\beta}dx^\alpha dx^\beta=-f(\phi,Q)dt^2+\frac{dx^2}{f(\phi,Q)}.
\end{equation}
Here, the metric function $f(\phi,Q)$ is given by
\begin{equation}
 f(\phi,Q)=J_Q(\phi,Q)-{\cal C},
\end{equation}
 where $J_Q(\phi,Q)$ is the integration of $V_Q$ given by
\begin{equation}
J_Q(\phi,Q)=\int^{\phi}V_Q(\tilde{\phi},Q)d\tilde{\phi}=\sqrt{\phi}+\frac{Q^2}{4\sqrt{\phi}}+\frac{4\phi\sqrt{\phi}}{l^2}.
\end{equation}
Also, ${\cal C}$ is a coordinate-invariant constant of
integration, which is identified with the mass $M_Q$ of the RN-AdS
black hole. We note here an  important connection between
$J_Q(\phi)$ and the metric function $U(\phi)$:
$J_Q(\phi)-M=\sqrt{\phi}U(\phi)$. Then, we recover the $t$-$r$ line
element for RN-AdS black hole from the 2D dilaton gravity approach
with $\phi=x=r^2/4$
\begin{equation}
 ds^2_{2D}=\sqrt{\phi}h_{\alpha\beta}dx^\alpha dx^\beta=\sqrt{\phi}\Bigg[-U(r)dt^2+\frac{dr^2}{U(r)}\Bigg]=
 \frac{r}{2}ds^2_{RN-AdS}.
\end{equation}
We note that the outer and inner  horizons appear when satisfying
\begin{equation}
f(\phi_\pm)=0 \to U(r_\pm)=0.
\end{equation}
In this case we find from Eq. (\ref{phisq}) that $(\nabla\phi)^2=0$,
even for $\phi'\not=0$. Note that if one starts from Eq.
(\ref{non-conf}) without the conformal transformation, it is
tedious to find the solution of $h_{\alpha\beta}={\rm
diag}[-U(r),U(r)^{-1}]$ by solving the full Einstein and dilation
equations. Hence the conformal transformation of Eq. (\ref{conft})
provides an efficient way to find the 4D black hole solution when
using  the 2D dilaton gravity. Further, we could easily recover 4D
thermodynamic quantities when using the conformal transformation.
Given that the considering 4D model  admits black hole solution, it
can be shown that all thermodynamic quantities of the black hole except the free energy are
invariant under the conformal transformations.

\begin{figure}[t!]
   \centering
   \includegraphics{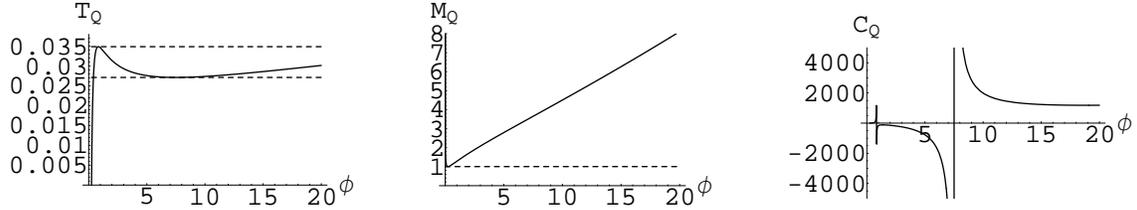}
\caption{Three graphs for the 2D dilaton gravity with $Q=1$.
(a) The solid curve describes $T_Q$ with two dashed horizons
for $T_Q = T_D(=0.035),~T_0(=0.027)$ at
$\phi = \phi_D(=0.83),~\phi_0(=7.5)$, respectively. (b) Mass $M_Q$ whose
minimum is $M_e=1.005$ at $\phi=\phi_e(=0.24)$. (c) Heat
capacity $C_Q$ which diverges at $\phi=\phi_D,~\phi_0$. }
\label{fig2}
\end{figure}
All thermodynamic quantities can be explicitly expressed
in terms of  the dilaton $\phi$, the dilaton  potential $V_Q(\phi,Q)$,
its integration $J_Q(\phi,Q)$, and its derivative
$V_Q'(\phi,Q)$ with respect to $\phi$ as
 \begin{equation} \label{phitds}
 S=4\pi \phi,~M_Q(\phi,Q)=J_Q(\phi,Q),~T_Q(\phi,Q)=\frac{V_Q(\phi,Q)}{4\pi},
 ~C_Q(\phi,Q) = 4\pi \frac{V_Q(\phi,Q)}{V_Q'(\phi,Q)}.
\end{equation}

In Fig. 2, we have the corresponding dual graphs, which are nearly
the same as in Fig. 1. At the extremal point $\phi=\phi_e$, we have $T_Q=0$ and $C_Q=0$,
which are determined by $V_Q(\phi_e)=0$. On the other hand, at the
local maximum point $\phi=\phi_D$, one has $T_Q=T_D,~C_Q=\pm \infty$,
which are determined  by the condition of $V_Q'(\phi_D)=0$.
Finally, at the local minimum point $\phi=\phi_0$, one has
$T_Q=T_0,~C_Q=\pm \infty$, which are determined  by the condition
of $V_Q'(\phi_0)=0$.  For $\phi_e<\phi<\phi_D$, we have the near-horizon phase of  extremal black hole,
whereas for $\phi>\phi_D$, we have the Schwarzschild-AdS
phase.

On the other hand, for the fixed-potential ensemble, we have a slight different form
\begin{equation}
V_\Phi(\phi,\Phi)=\frac{1}{2\sqrt{\phi}}\Big[1-\Phi^2\Big]
+\frac{6\sqrt{\phi}}{l^2}. \label{potp}
\end{equation}
Its derivative with respect to $\phi$ takes the form
\begin{equation}
\frac{dV_\Phi(\phi,\Phi)}{d\phi}=\frac{1}{4\phi^{3/2}}\Big[-1+\Phi^2\Big]+\frac{3}{l^2\sqrt{\phi}}.
\end{equation}
Its integration over $\phi$ leads to the internal energy
\begin{equation}
J_\Phi(\phi,\Phi)=\sqrt{\phi}\Big[1-\Phi^2\Big]+\frac{4\phi\sqrt{\phi}}{l^2}.
\end{equation}
However, the ADM mass is given by
\begin{equation}
M_\Phi(\phi,\Phi)=J_\Phi+\Phi
Q=\sqrt{\phi}\Big[1+\Phi^2\Big]+\frac{4\phi\sqrt{\phi}}{l^2}.
\end{equation}
Other thermodynamic quantities are shown to have
\begin{equation} \label{phithp}
 T_\Phi(\phi,\Phi)=\frac{V_\Phi(\phi,\Phi)}{4\pi},~C_\Phi(\phi,\Phi) =
 T_\Phi\Big(\frac{\partial S}{\partial T_\Phi}\Big)_\Phi= 4\pi \frac{V_\Phi(\phi,\Phi)}{V_\Phi'(\phi,\Phi)}.
\end{equation}
For $\Phi<\Phi_c=1$, their behaviors are depicted in Fig. 3. In
this case, we have no degenerate horizon but a blow-up  heat
capacity at $\phi=\tilde{\phi}_0$. On the other hand, for $\Phi\ge
\Phi_c=1$, we have degenerate horizon which satisfies $V_\Phi=0$
but not a blow-up heat capacity. Also, Eq. (\ref{newat2}) takes
the form
\begin{equation}\label{newat2p}
\bar{R}_\Phi=-\frac{dV_\Phi(\phi,\Phi)}{d\phi}.
\end{equation}
\begin{figure}[t!]
   \centering
   \includegraphics{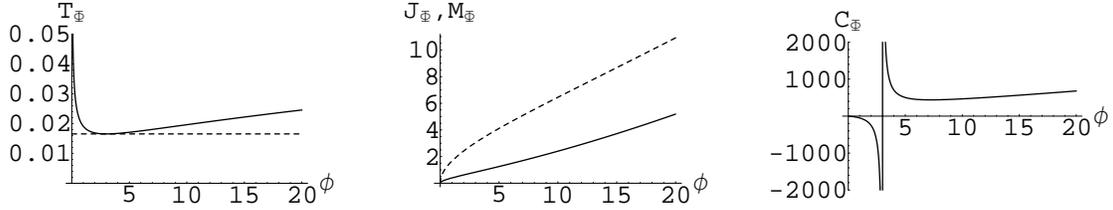}
\caption{Three graphs for the 2D dilaton gravity with
$\Phi=0.8<\Phi_c$. (a) The solid curve describes $T_\Phi$ with
dashed horizon for the minimum temperature $T=T_0(=0.017)$ at
$\phi=\tilde{\phi}_0(=3)$. (b) Internal energy $J_\Phi$ (solid
curve) and mass $M_\Phi$ (dashed curve). (c) Heat capacity
$C_\Phi$ which diverges at $\phi=\tilde{\phi}_0$.} \label{fig3}
\end{figure}

\section{ thermodynamic relations}

\subsection{Fixed-charge ensemble}
First, let us consider the fixed-charge $Q$  ensemble.  Using the
relation of $r_+=\sqrt{S/\pi}$ together with Eqs. (\ref{aas}) and
(\ref{aac}) or $\phi=S/4\pi$ with Eq. (\ref{phitds}), we can
rewrite thermodynamic quantities as functions of the entropy $S$
and charge $Q$
\begin{eqnarray}\label{ats}
M_Q(S,Q)&=&\frac{1}{2}\sqrt{\frac{S}{\pi}}\Bigg(1+ \frac{\pi Q^2}{S}+\frac{S}{\pi l^2}\Bigg)\\
T_Q(S,Q)&=&\frac{1}{4\sqrt{\pi S}}\Bigg(1-\frac{\pi Q^2}{S} +\frac{3S}{\pi l^2}\Bigg),\\
C_Q(S,Q)&=& 2S\Bigg[ \frac{1-\frac{\pi Q^2}{S}+\frac{3S}{\pi
l^2}}{-1+\frac{3\pi Q^2}{S}+\frac{3S}{\pi l^2}}\Bigg].
\end{eqnarray}
We introduce the first-law of the thermodynamics
\begin{equation} \label{FLT}
dM_Q = T_Q dS + \Phi dQ
\end{equation}
with the chemical potential $\Phi$.
We have the thermodynamic relations
\begin{equation}
T_Q=\Big(\frac{\partial M_Q}{\partial S}\Big)_Q,~
\Big(\frac{\partial T_Q}{\partial S}\Big)_Q=\Big(\frac{\partial^2
M_Q}{\partial S^2}\Big)_Q.
\end{equation}
From Eqs.(\ref{newat2}) and (\ref{derivv}),  the 2D curvature
scalar $\bar{R}_Q$ defined by Eq.(\ref{2dcur}) takes the form
\begin{equation}\label{2dcurq}
\bar{R}_Q(S,Q)=-(4 \pi)^2 \Big(\frac{\partial T_Q}{\partial
S}\Big)_Q= -\frac{2 \pi^{3/2}}{S\sqrt{S}}\Bigg(-1+\frac{3\pi
Q^2}{S}+\frac{3S}{\pi l^2}\Bigg),
\end{equation}
which shows that the 2D curvature constructed from
Eq.(\ref{2dmetric}) could describe thermodynamics of RN-AdS black
hole when using the 2D Einstein equation. Importantly, the heat
capacity can be rewritten as
\begin{equation}
C_Q(S,Q)=T_Q\Big(\frac{\partial T_Q}{\partial S}\Big)^{-1}_Q=-(4
\pi)^2\frac{T_Q}{\bar{R}_Q},
\end{equation}
which shows clearly that $C_Q$ blows up when $\bar{R}_Q=0$, and it
is zero when $T_Q=0$.
As is shown in Fig. 4, the 2D curvature  shows that $\bar{R}_Q=-4.42$ (AdS$_2$ spacetime) at the extremal
point of $\phi=\phi_e$, while it is zero (flat spacetime) at the Davies' point $\phi=\phi_D$ and minimum temperature point
$\phi=\phi_0$. This shows a close connection between the heat capacity and 2D curvature in the canonical ensemble.

We note one  connection such as
\begin{equation}
1-\frac{\pi Q^2}{S} +\frac{3S}{\pi l^2}=0 \longleftrightarrow
l^2r_+^2-l^2Q^2+3r_+^4=0
\end{equation}
whose solution is $r_+^2=r_e^2,~S=S_e=\pi r_e^2$.
We also have another connection,
\begin{equation}
-1+\frac{3\pi Q^2}{S} +\frac{3S}{\pi l^2}=0 \longleftrightarrow
-l^2r_+^2+3l^2Q^2+3r_+^4=0
\end{equation}
whose solution is $r^2_+=r_D^2,~S=S_D=\pi r_D^2$ and
$r^2_+=r_0^2,~S=S_0=\pi r_0^2$.
\begin{figure*}[t!]
   \centering
   \includegraphics{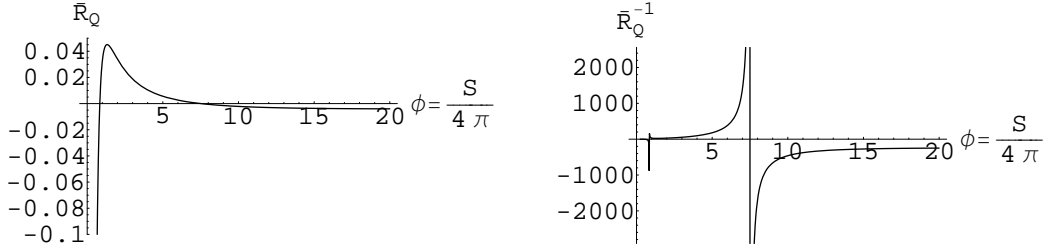}
\caption{Graphs for 2D curvature $\bar{R}_Q$ and its inverse
curvature $\bar{R}^{-1}_Q$ as functions of $\phi=S/4\pi$ for
$Q=1<Q_c$ . We find that $\bar{R}_Q=-4.42,0,0$ for $\phi=\phi_e,
\phi_D,\phi_0$, respectively. We have AdS$_2$,~flat,~dS$_2$
spacetimes, depending on the value of $\bar{R}_Q<0,~=0,~>0$. On
the other hand, $\bar{R}^{-1}_Q$ blows up at $\phi=
\phi_D,\phi_0$, showing similar behavior of heat capacity $C_Q$. }
\label{fig4}
\end{figure*}
\subsection{Fixed-potential ensemble}

Next, let us consider the fixed-potential
ensemble. We can rewrite thermodynamic quantities as functions of
the entropy $S$ and the chemical potential $\Phi$
\begin{eqnarray}\label{atsp}
J_\Phi(S,\Phi)&=&M_\Phi-\Phi Q=\frac{1}{2}\sqrt{\frac{S}{\pi}}\Bigg(1-\Phi^2+\frac{S}{\pi l^2}\Bigg),\\
T_\Phi(S,\Phi)&=&\frac{1}{4\sqrt{\pi S}}\Bigg(1-\Phi^2 +\frac{3S}{\pi l^2}\Bigg),\\
C_\Phi(S,\Phi)&=& 2S\Bigg[ \frac{1-\Phi^2+\frac{3S}{\pi
l^2}}{-1+\Phi^2+\frac{3S}{\pi l^2}}\Bigg],
\end{eqnarray}
where
\begin{equation}
M_\Phi(S,\Phi)= \frac{1}{2}\sqrt{\frac{S}{\pi}}\Bigg(1+\Phi^2+\frac{S}{\pi l^2}\Bigg).
\end{equation}
For $\Phi>\Phi_c=1$, we could find zero temperature which implies
the presence of extremal black hole at $S=S_e=\frac{\pi
l^2}{3}[\Phi^2-1]$. Also, in the case of $\Phi=\Phi_c$, we could
find zero temperature at $S=0$. For $\Phi<\Phi_c$, we could not
find zero temperature for any $S$. However, we find that the heat
capacity blows up for $\Phi<\Phi_c$ at $S=S_0=\frac{\pi
l^2}{3}[1-\Phi^2]$. In this case, we expect that the RN-AdS black
hole shows a similar behavior as the Schwarzschild-AdS black hole.

Now, let us introduce the law of the thermodynamics for the
$J_\Phi(S,\Phi)$
\begin{equation} \label{FLTP}
dJ_\Phi=T_\Phi dS-Qd\Phi
\end{equation}
with $dM_\Phi=T_\Phi dS+Q d\Phi$. Also note that the redefined
mass $J_\Phi$ is identified with a free energy of the black hole
because the Gibbs free energy is defined as $G=J_\Phi-T_\Phi S$.
We note that the second term represents the work done by the black
hole on its environment. Then, for fixed-charge $\Phi$ ensemble,
we have the thermodynamic relations
\begin{equation}
T_\Phi=\Big(\frac{\partial J_\Phi}{\partial S}\Big)_\Phi,~
\Big(\frac{\partial T_\Phi}{\partial
S}\Big)_\Phi=\Big(\frac{\partial^2 J_\Phi}{\partial
S^2}\Big)_\Phi.
\end{equation}

From Eq. (\ref{newat2}), the 2D curvature scalar $\bar{R}_\Phi$
takes the form
\begin{equation} \label{2dcurp}
\bar{R}_\Phi(S,\Phi)=-(4\pi)^2\Big(\frac{\partial^2
J_\Phi}{\partial S^2}\Big)_\Phi= -\frac{2
\pi^{3/2}}{S\sqrt{S}}\Bigg(-1+\Phi^2+\frac{3S}{\pi l^2}\Bigg).
\end{equation}
On the other hand, the heat capacity can be rewritten as
\begin{equation}
C_\Phi(S,\Phi)=T_\Phi\Big(\frac{\partial T_\Phi}{\partial
S}\Big)^{-1}_\Phi=-(4 \pi)^2\frac{T_\Phi}{\bar{R}_\Phi},
\end{equation}
which shows clearly that $C_\Phi$ blows up when $\bar{R}_\Phi=0$.
As is shown in Fig. 5, the 2D curvature  shows that  it is zero
(flat spacetime) at the minimum temperature point
$\phi=\tilde{\phi}_0$. Similar to the canonical ensemble, this
also shows a close connection between the heat capacity and 2D
curvature in the grand canonical ensemble.
\begin{figure*}[t!]
   \centering
   \includegraphics{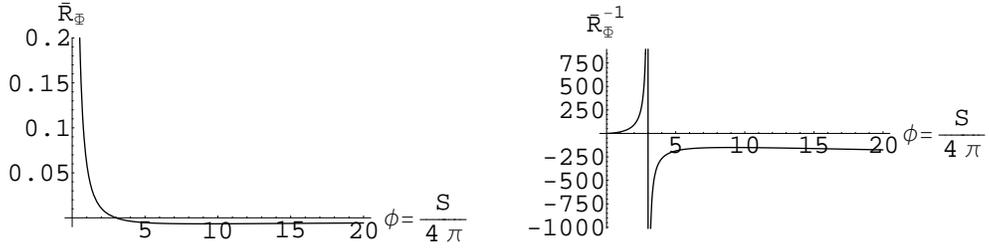}
\caption{Graphs for 2D curvature $\bar{R}_\Phi$ and its inverse
curvature  $\bar{R}^{-1}_\Phi$ as functions of $\phi=S/4\pi$ for
$\Phi=0.8<\Phi_c$. We find that $\bar{R}_\Phi=0$ for
$\phi=\tilde{\phi}_0$. We have dS$_2$, flat, and AdS$_2$
spacetimes, depending on the value of $\bar{R}_\Phi>0,~=0,~<0$,
respectively. On the other hand, $\bar{R}^{-1}_\Phi$ blows up at
$\phi=\tilde{\phi}_0$, showing similar behavior of heat capacity
$C_\Phi$. } \label{fig5}
\end{figure*}
\section{ Discussions}

We summarize the features of heat capacity, Ruppeiner curvature,
and  inverse 2D curvature at the extremal, Davies' and minimum
temperature points in the $(S,Q)$-representation in Table I. We do
not find any consistent relations between the heat capacity and
four Ruppeiner curvatures. We note that ${\cal R} \propto 1/({\rm
Det}[g_{ij}])^2$. If the Ruppeiner metric is not on-diagonal form,
like Eq. (\ref{metQ}), then the pole of  ${\cal R}$ is shifted
from $1/(g_{SS}g_{QQ})^2=(C_Q/g_{QQ})^2  \to
1/(g_{SS}g_{QQ}-g_{SQ}^2)^2$: $(1-\frac{3\pi Q^2}{S}-\frac{3S}{\pi
l^2})^2=0 \to  (1-\frac{\pi Q^2}{S}-\frac{3S}{\pi l^2})^2=0$.
Hence the Ruppeiner curvature does not show the feature of heat
capacity at the Davies' and minimum temperature points where heat
capacity blows up. However, if the metric is on-diagonal
($g_{SQ}=0$), the Ruppeiner curvature preserves the pole of heat
capacity as  $1/g^2_{SS}=C^2_Q$ is shown.

In the $(S,\Phi)$-representation, the same happens. Since its
Ruppeiner metric of Eq. (\ref{metP}) is not on-diagonal form, the
pole of ${\cal R}$ is shifted from
$1/(g_{SS}g_{\Phi\Phi})^2=(C_\Phi/g_{\Phi\Phi})^2 \to
1/(g_{SS}g_{\Phi\Phi}-g_{S\Phi}^2)^2$: $(1-\Phi^2-\frac{3S}{\pi
l^2})^2=0 \to  (1-3\Phi^2-\frac{3S}{\pi l^2})^2=0$. Hence the
Ruppeiner curvature does not preserve the feature of heat capacity
at minimum temperature point where heat capacity blows up.

At this stage, we comment on the connection between Ruppeiner
geometric and thermodynamic ensemble approaches. In the canonical
ensemble of fixed-charge, we have $g^Q_{ij}={\rm
diag}[g^Q_{SS},0]$, while in the grand canonical ensemble of
fixed-potential, we have $g^\Phi_{ij}={\rm diag}[g^\Phi_{SS},0]$.
A single component of $g^{Q/\Phi}_{SS}$ contains  the whole
information on heat capacity of RN-AdS black hole. These two cases
lead to flat thermodynamic space  of ${\cal R}^{Q/\Phi}=0$. It
seems that Ruppeiner geometric approach is not compatible with the
conventional thermodynamic ensemble approach of black holes.

Although we have performed for the RN-AdS black hole, this
behavior persists to any black holes. We have shown that all
thermodynamic quantities of the RN-AdS black holes can be
described from the 2D dilaton gravity obtained by performing the
dimensional reduction and conformal transformation. Also it is
known that entropy, temperature, mass and specific heat are
invariant under the conformal transformations.

Hence, as a consistent connection to the membrane paradigm of
black holes, we  propose the promising 2D curvature, which carries
out the information on the heat capacity. The inverse 2D curvature
could well describe Davies' and minimum temperature points, as the
heat capacity does.

Finally,  considering Eqs.(\ref{2dcurq}) and(\ref{2dcurp}), we
confirm that the negativity of $\bar{R}$(positivity of
$\partial^2M/\partial S^2$) is sufficient to ensure the local
stability of RN-AdS black holes~\cite{CGG}.

\begin{table*}
\begin{tabular}{|c|c|c|c|c|}
  \hline
   & Extremal point(RN)  & Davies' point(RN)  & Minimum point(SAdS) & Remarks \\
  \hline
   $C_Q$ & 0 & diverges & diverges & Heat capacity
   \\\hline
   ${\cal R}^I_{Q}$ & diverges & finite & finite & RC
   \\\hline
   ${\cal R}^{II}_{Q}$& diverges & finite & finite & RC
   \\\hline
  $\bar{R}_2^{-1}$ & finite & diverges & diverges & 2D curvature\\
  \hline
\end{tabular}

\label{table I}
  \caption{ Heat capacity $C_Q$, Ruppeiner curvatures ${\cal R}^{I}_Q$ in\cite{abp} and
   ${\cal R}^{II}_Q$ in\cite{mz}, and inverse 2D curvature $\bar{R}^{-1}_Q$ for extremal,
  Davies, and minimum temperature points of RN-AdS black hole for $Q<Q_c$.
  RN (SAdS) denote the Reissner-Norstr\"om (Schwarzschild-AdS) black holes.}
\end{table*}

\section*{Acknowledgement}

The authors thank Joey Medved for his helpful discussions.
Two of us (Y.S. Myung and Y.-J. Park) were supported by the
Science Research Center Program of the Korea Science and
Engineering Foundation through the Center for Quantum Spacetime of
Sogang University with grant number R11-2005-021.  Y.-W. Kim was
supported by the Korea Research Foundation Grant funded by Korea
Government (MOEHRD): KRF-2007-359-C00007.


\begin{thebibliography}{99}


\bibitem{bek} J. D. Bekenstein, Lett. Nuovo Cimento 4 (1972) 737.
\bibitem{bek2} J. D. Bekenstein, Phys. Rev. D 7 (1973) 2333.
\bibitem{bek3} J. D. Bekenstein, Phys. Rev. D 9 (1974) 3292.
\bibitem{haw} S. W. Hawking, Commun. Math. Phys. 43 (1975) 199.
\bibitem{BCH}   J.~M.~Bardeen, B.~Carter and S.~W.~Hawking,
            Commun. Math. Phys. 31 (1973) 161.

\bibitem{wr} F. Weinhold, J. Chem. Phys. 63 (1975) 2479;\\
             G. Ruppeiner, Phys. Rev. A 20 (1979) 1608;\\
             G. Ruppeiner, Rev. Mod. Phys. 67 (1995) 605;
              68 (1996) 313(E).
\bibitem{dav} P. C. W. Davies, Proc. R. Soc. Lond. A 353 (1977) 499.
\bibitem{rup} S. Ferrara, G. W. Gibbons, and R. Kallosh,
              Nuc. Phys. B 500 (1997) 75;\\
              R.~G.~Cai and J.~H.~Cho,
              Phys. Rev. D 60 (1999) 067502;\\
              G. Arcioni and E. Lozano-Tellechea, Phys. Rev. D 72 (2005)
              104021;\\
J. Aman, and N. Pidokrajt, Phys. Rev. D 73 (2006) 024017;\\
J. Aman, I. Bengtsson and N. Pidokrajt, Gen. Rel. Grav. 38 (2006)
1305;\\
H.~Quevedo,
  J.\ Math.\ Phys.\  {\bf 48} (2007) 013506
  [arXiv:physics/0604164];

T. Sarkar, G. Sengupta and B. N. Tiwari, JHEP 0611 (2006) 015;\\
J.~L.~Alvarez, H.~Quevedo and A.~Sanchez,
  arXiv:0801.2279 [gr-qc];\\
A.J.M. Medved, ``A Commentary on Ruppeiner Metrics for Black
Holes", arXiv:0801.3497 [gr-qc].
\bibitem{abp} J.~Aman, I.~Bengtsson and N.~Pidokrajt,
  Gen. Rel. Grav. 35 (2003) 1733.

  \bibitem{que} H.~Quevedo,
  ``Geometrothermodynamics of black holes,''
  arXiv:0704.3102 [gr-qc].

\bibitem{SCWS}J.~Shen, R.~G.~Cai, B.~Wang and R.~K.~Su,
  Int. J. Mod. Phys. A {\bf 22}, 11 (2007).


\bibitem{mz} B.~Mirza and M.~Zamani-Nasab,
  JHEP 0706 (2007) 059.

   \bibitem{wein} F. Weinhold, Phys. Today {\bf 29} (1976) 23.


\bibitem{Rup1} G. Ruppeiner,
``Thermodynamic curvature and phase transitions in Kerr-Newman
black holes'', arXiv:0802.1326 [gr-qc].

  \bibitem{Rup2} G. Ruppeiner, ``Black Holes: Fermions at the Extremal Limit?'',
arXiv:0711.4328 [gr-qc]

\bibitem{Tho} K. S. Thorne, R. H. Price, D. A. Macdonald, {\it Black Holes: The Mem-
brane Paradigm} (Yale University Press, New Haven, 1986).

\bibitem{witten} E. Witten, Phys. Rev. D 44 (1991) 314.
\bibitem{wit1} G. Mandal, A. M. Sengupta, S. R. Wadia,
                Mod. Phys. Lett. A 6 (1991) 1685.

\bibitem{crff} C. G. Callan, S. B. Giddings, J. A. Harvey,
                 A. Strominger, Phys. Rev. D 45 (1992) R1005.
\bibitem{crf1} J. G. Russo, L. Susskind, L. Thorlacius,
                Phys. Lett. B 292 (1992) 13.
\bibitem{crf2} V. P. Frolov, Phys. Rev. D 46 (1992) 5383.

\bibitem{gkv} D. Grumiller, W. Kummer, D. V. Vassilevich,
Phys. Rept. 369 (2002) 327.

\bibitem{jackiw} R. Jackiw,  in {\it Quantum Theory of Gravity},
                ed. S. M. Christensen (Hilger, Bristol, 1984).
\bibitem{Teit} C. Teitelboim, in {\it Quantum Theory of Gravity},
                 ed. S. M. Christensen (Hilger, Bristol, 1984).

\bibitem{JT-theories} M. Henneaux, Phys. Rev. Lett. 54 (1985) 959.

\bibitem{JT2} R. B. Mann, D. Robbins, and T. Ohta,
Phys. Rev. Lett. 82 (1999) 3738.

\bibitem{fnn} A. Fabbri, D. J. Navarro, and J. Navarro-Salas,
Nucl. Phys. B 595 (2001) 381.

\bibitem{mod} Y.S. Myung, Y.-W. Kim and Y.-J. Park,
Mod. Phys. Lett. A 23 (2008) 91;\\
Y.S. Myung, Y.-W. Kim and Y.-J. Park, ``Thermodynamics of regular
black hole'', arXiv:0708.3145 [gr-qc];\\
Y.S. Myung, ``Phase transition between non-extremal and extremal
Reissner-Nordstrom black holes'', arXiv:0710.2568 [gr-qc].
\bibitem{CEJM}
  A.~Chamblin, R.~Emparan, C.~V.~Johnson and R.~C.~Myers,
  Phys. Rev. D 60 (1999) 064018.

  \bibitem{TL} T. Tranah and P. T. Landsberg, Collective Phenomena
{\bf 3} (1980) 81

  \bibitem{Wu}
  X.~N.~Wu,
  Phys. Rev. D 62 (2000) 124023.

\bibitem{great}
  General Relativity, Einstein \& All package (GREAT.m) (2003),\\
   http://library.wolfram.com/infocenter/MathSource/4781/

\bibitem{GKL}
  J.~Gegenberg, G.~Kunstatter and D.~Louis-Martinez,
  Phys. Rev. D 51 (1995) 1781.

\bibitem{cfnn} J.~Cruz, A.~Fabbri, D.~J.~Navarro, and J.~Navarro-Salas,
          Phys. Rev. D 61 (2000) 024011.
 \bibitem{mkp}
  Y.~S.~Myung, Y.~W.~Kim and Y.~J.~Park,
  Phys. Rev. D 76 (2007) 104045.

  \bibitem{GM} D.~Grumiller and R.~McNees,
  JHEP  0704  (2007) 074.

\bibitem{NSN}
  J.~Navarro-Salas and P.~Navarro,
  Nucl. Phys. B   579 (2000) 250.

  \bibitem{CGG}
  M.~Cvetic and S.~S.~Gubser,
  JHEP {\bf 9904} (1999) 024
  [arXiv:hep-th/9902195];
  S.~S.~Gubser and I.~Mitra,
  JHEP {\bf 0108} (2001) 018
  [arXiv:hep-th/0011127];
A.~Sheykhi and N.~Riazi,
  Phys.\ Rev.\  D {\bf 75} (2007) 024021
  [arXiv:hep-th/0610085].





\end{thebibliography}
\end{document}